\documentclass[superscriptaddress,nofootinbib]{revtex4}

\usepackage{graphicx}
\usepackage{axodraw}
\usepackage{epsfig}
\newcommand{\tplus}{T^+}
\newcommand{\tminus}{T^-}
\newcommand{\tpm}{T^{\pm}}
\newcommand{\tzero}{T^0}
\newcommand{\tzerobar}{\bar{T}^0}
\newcommand{\missinget}{\slash\hspace{-0.1 in}{E}_T}

\begin{document}
\bibliographystyle{revtex}


\title{Phenomenology of a Constrained Standard Model from an Extra Dimension}



\author{M. Chertok}
\affiliation{University of California, Davis, CA 95616}

\author{G.D. Kribs}
\affiliation{University of Wisconsin, Madison,  WI 53706}

\author{Y. Nomura}
\affiliation{University of California, Berkeley, CA 94720}

\author{W. Orejudos}
\affiliation{Lawrence Berkeley National Laboratory, Berkeley, CA 94720}

\author{B. Schumm}
\affiliation{Santa Cruz Institute for Particle Physics and the University of 
California, Santa Cruz, CA 95064}

\author{S. Su}
\affiliation{California Institute of Technology, Pasadena, CA 91125}



\begin{abstract}
We describe a highly predictive model for supersymmetry 
breaking in 5 dimensions.  
We develop its phenomenology and the capabilities for discovery and 
measurement at various colliders.
\end{abstract}

\maketitle


\section{Introduction and theoretical motivation}
\label{P3.10intro}

Supersymmetry (SUSY) is the leading candidate for physics beyond the
Standard Model (SM) because it stabilizes the Higgs mass against 
quadratically divergent radiative corrections. 
Since no evidence for superpartners has been found, 
supersymmetry cannot be exact.  Breaking supersymmetry in an 
experimentally viable way is therefore the premiere stumbling block 
towards constructing realistic low energy supersymmetric models.

The usual approach to breaking supersymmetry is to write a
low energy effective theory, the minimal supersymmetric standard
model (MSSM), with mass and coupling terms that
break supersymmetry ``softly''.  
This soft breaking introduces over a hundred new parameters with no 
\emph{a priori} guidance as to their size.  The vast majority of this 
parameter space is excluded by stringent experimental constraints
on flavor-changing neutral current processes.
Furthermore, an extra Higgs doublet (supermultiplet) 
is necessary for consistency of the model, allowing the
supersymmetry-preserving mass term $\mu H_u H_d$ in the superpotential.
Electroweak symmetry breaking requires $\mu$ of order the electroweak 
scale, and there is no understanding why this \emph{supersymmetric}
mass term should be of the same order as the soft supersymmetry 
\emph{breaking} mass terms.  Thus, there are strong motivations to look 
for an organizing principle associated with supersymmetry breaking
that can naturally explain the size and pattern of superpartner masses.

\section{Overview of model}
\label{P3.10theory}
Recently, a new approach to low energy supersymmetry has been
given by Barbieri, Hall, and Nomura \cite{BHN}.  Unlike the usual
approach of postulating a low energy effective theory with so-called
``soft'' supersymmetry breaking terms added by hand, the entire SM 
is supersymmetrized in \emph{five} dimensions. 
This means there are not only complete (${\cal N}=1$ in 4D) supermultiplets 
consisting of the SM fields and their superpartners, but also a 
``mirror'' set of supermultiplets consisting of mirror SM fields
and their mirror superpartners.  This is required since supersymmetry
in 5D has double the number of supercharges than in 4D; i.e., 
an ${\cal N}=1$ supermultiplet in 5D has the field content of
a single ${\cal N}=2$ supermultiplet in 4D, which is equivalent
to \emph{two} ${\cal N}=1$ supermultiplets in 4D.

\begin{figure}
\begin{center} 
\begin{picture}(350,90)(-10,-10)

  \Line(5,0)(320,0)
  \LongArrow(10,-10)(10,80)
  \Text(-40,30)[b]{mass}
  \Text(0,0)[r]{$0$}
  \Line(8,20)(12,20)    \Text(6,20)[r]{$1/R$}
  \Line(8,40)(12,40)    \Text(6,40)[r]{$2/R$}
  \Line(8,60)(12,60)    \Text(6,60)[r]{$3/R$}
  \Text(60,80)[b]{$\psi_{M}, \phi_{H}, A^{\mu}$}
  \Line(40,0)(80,0)      \Vertex(60,0){3}

  \Line(40,40)(80,40)    \Vertex(60,40){3}
  \Text(130,80)[b]{$\phi_{M}, \psi_{H}, \lambda$}

  \Line(110,20)(150,20)    \Vertex(130,20){3}
  \Line(110,60)(150,60)    \Vertex(130,60){3}
  \Text(200,80)[b]{$\phi^{c}_{M}, \psi^{c}_{H}, \psi_{\Sigma}$}

  \Line(180,20)(220,20)    \Vertex(200,20){3}
  \Line(180,60)(220,60)    \Vertex(200,60){3}
  \Text(270,80)[b]{$\psi^{c}_{M}, \phi^{c}_{H}, \phi_{\Sigma}$}
  \Line(250,40)(290,40)    \Vertex(270,40){3}


  \Text(35,7)[r]{$h$} \Text(40,7)[l]{- - - $\ $- - - - } \Vertex(60,7){3}
  \Text(105,10)[r]{$\tilde{t}_1$} \Text(110,10)[l]{- - - $\ $- - - - } 
 \Vertex(130,10){3} 
  \Text(105,30)[r]{$\tilde{t}_2$} \Text(110,30)[l]{- - - $\ $- - - - } 
\Vertex(130,30){3} 
  \Text(175,10)[r]{$\tilde{t}_1^c$}  \Text(180,10)[l]{- - - $\ $- - - - } 
\Vertex(200,10){3} 
  \Text(175,30)[r]{$\tilde{t}_2^c$}  \Text(180,30)[l]{- - - $\ $- - - - } 
 \Vertex(200,30){3} 
\end{picture}

\caption{Tree-level KK mass spectrum of the matter, Higgs and gauge 
multiplets.  Physical light Higgs and top squarks mass eigenstates are 
shown in dashed lines. }
\label{P3.10Fig_spectrum}
\end{center}
\end{figure}
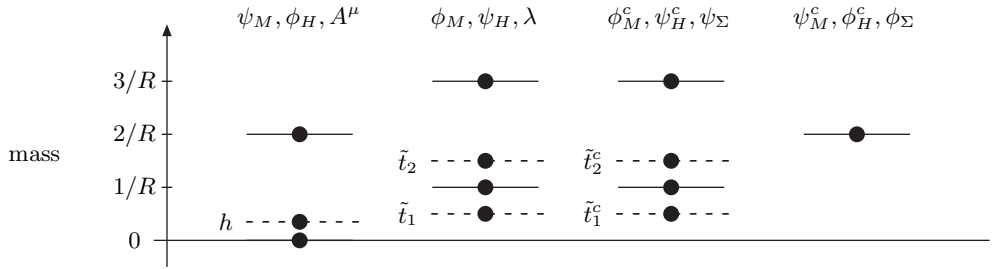

The 5D spacetime is assumed to be compactified on an $S_1/(Z_2 \times Z_2')$
orbifold.  Thus, the physical space is a line segment with
two ends -- the orbifold fixed points.  At each fixed point the 5D 
fields can transform as either even or odd under the $Z_2$ symmetry
associated with that fixed point, as shown in Fig.~\ref{P3.10Fig_transf}.
\begin{figure}
\begin{center} 
\begin{picture}(450,30)(-260,0)
  \Text(-200,25)[b]{$\psi_{M} (+,+)$}
  \Text(-215,15)[r]{$\phi_{M} (+,-)$}
  \Text(-185,15)[l]{$\phi^{c\dagger}_{M} (-,+)$}
  \Text(-200,5)[t]{$\psi^{c\dagger}_{M} (-,-)$}
  \Text(-40,25)[b]{$\psi_{H} (+,-)$}
  \Text(-55,15)[r]{$\phi_{H} (+,+)$}
  \Text(-25,15)[l]{$\phi^{c\dagger}_{H} (-,-)$}
  \Text(-40,5)[t]{$\psi^{c\dagger}_{H} (-,+)$}
  \Text(120,25)[b]{$A^{\mu} (+,+)$}
  \Text(105,15)[r]{$\lambda (+,-)$}
  \Text(135,15)[l]{$\psi_{\Sigma} (-,+)$}
  \Text(120,5)[t]{$\phi_{\Sigma} (-,-)$}
\end{picture}
\caption{Quantum numbers of the matter, Higgs and gauge 
multiplets under the two orbifoldings:
($\psi_M$, $\phi_M$, $\psi_M^c$, $\phi_M^c$) 
are the matter 
(fermions, sfermions, mirror fermions, mirror sfermions),
($\psi_H$, $\phi_H$, $\psi_H^c$, $\phi_H^c$) 
are the 
(Higgsinos, Higgs scalars, mirror Higgsinos, mirror Higgs scalars),
and ($A^\mu$, $\lambda$, $\psi_\Sigma$, $\phi_\Sigma$) 
are the 
(gauge bosons, gauginos, adjoint fermions, and adjoint scalars).
The ($\pm$,$\pm$) labels refer to the ($Z_2,Z_2'$)
parity properties, and thus determine the allowed wave functions of 
these fields.  
}
\label{P3.10Fig_transf}
\end{center}
\end{figure}
The field content can be readily recognized as that of massless
${\cal N}=2$, 4D hypermultiplets and
vector multiplets.  The Kaluza-Klein (KK) reduction of this theory to 4D
yields wave functions as sines and cosines of integer and half-integer
multiplets of the size of the extra dimension, $R$, with a
mass spectrum given in Fig.~\ref{P3.10Fig_spectrum} (solid lines).
Note that no supersymmetry is preserved across the entire dimension, 
and thus the zero modes consist \emph{only} of SM fields.
There is no $\mu$ term in this model, due to ${\cal N}=1$ SUSY in 5D.  
The lightest Higgsinos and one Higgs doublet are massive, 
since they are odd under one or both of the $Z_2$'s.

An interesting feature of this model is that the Higgs effective potential 
can be calculated essentially in terms of a single free parameter $R$, 
which comes from the loop contribution through the top Yukawa coupling.
The only scale in the model, $1/R$,  
can be determined by the minimization condition of the 
Higgs effective potential, which gives $1/R\sim{370}$ GeV \cite{BHN-th}. 
This allows us to predict the physical Higgs boson mass as well as the 
superparticle and KK tower masses.  The predicted Higgs mass is 
$m_h = (127 \pm 8)~{\rm GeV}$.
At the first excited level, there are two superparticles for each SM 
particle.  Their masses shift from $1/R$ by electroweak breaking 
effects, such as fermion masses.
The largest effect appears in the top squark sector, giving top squarks 
of masses $1/R \pm m_t\sim{210,540}$ GeV.  
The mass spectrum of the light Higgs and stops is
also shown in Fig.~\ref{P3.10Fig_spectrum} as dashed lines.
Although possible 
additional brane localized interactions could shift
the value of $1/R$ by as much as a few tens of percent, the Higgs and stop mass
prediction is much less sensitive to the UV physics.  The observation of the 
light stop, described below, 
and the measurement of its mass in the predicted range would 
be an unmistakable signature of this model.

\section{Discussion of phenomenology}
\label{P3.10pheno}

The low energy effective theory \emph{below} $\sim{2}/R$ consists of
one Higgs doublet and two superpartners for each SM particle. 
The light Higgs has SM-like Yukawa couplings and $WWh$,
$ZZh$ gauge couplings.  It can be produced at Tevatron Run II via the usual
associated production of $Wh$ or $Zh$, with the Higgs decays into $b\bar{b}$ 
or $\tau\bar\tau$.  At LHC, $gg\rightarrow{h}\rightarrow\gamma\gamma$ 
would be the discovery channel for the light Higgs.
However, the top Yukawa 
coupling could have a $O(10\%)$ deviation from the SM value
due to the fact that it is a brane-localized interaction, which 
could be measured in a TeV scale $e^+e^-$ collider via $t\bar{t}h$ 
associated production \cite{tthlc}.  

One characteristic feature of this model is that the two degenerate 
light stops
\footnote{In the following, we use the symbol $\tilde{t}_1$($\tilde{t}_2$) 
to denote both degenerate light (heavy) stops.}
 ($\tilde{t}_1$) with mass $m_{\tilde{t}_1}=1/R-m_t$ are the LSPs, 
and are stable\footnote{The cosmological constraints on an absolutely 
stable stop could be relaxed if we allow small 
$R$-parity violation.} 
if $R$-parity is exact.
Since the stop carries color charge, once produced, it captures (anti)quarks 
in the detector matter to form a super-hadron. 
The lowest mass states \cite{oleg} are $\tplus=\tilde{t}_1\bar{d}$, 
$\tzero=\tilde{t}_1\bar{u}$ and their charge conjugate states 
$\tminus$, $\tzerobar$. 
Both neutral and charged states are sufficiently stable inside the detector
since the $\beta$-decay of $\tplus$ into $\tzero$ is suppressed by the
small mass splitting between them.  
$\tpm$ appears as a stiff charged track with little hadron calorimeter 
activity, and hits the muon chamber.  
A heavy $\tpm$ can be distinguished from a muon
via large $dE/dx$ or time-of-flight. 
$\tzero$ and $\tzerobar$ could be identified as missing energies
as they traverse the detector with little interactions.   

The other supersymmetric particles (besides the heavier stop) 
are almost degenerate in mass: $\sim{1}/R$, 
with small mass splittings coming from 
electroweak corrections or additional unknown brane kinetic terms. 
All of them cascade decay into stop LSPs.  
Squarks mostly decay via the channel
\begin{equation}
\tilde{q}\rightarrow{q}\tilde{g}\rightarrow{q}\tilde{t}_1{t}
\rightarrow{q}\tilde{t}_1{b}W^{\pm}\rightarrow{q}\tilde{t}_1{b}(l\nu/jj'),
\label{eq:P3.10squarkdecay}
\end{equation}
where $\tilde{g}$ in the first step can be either on-shell or off-shell. 
The $b$-jet from the top decay and the leptons/jets from the $W$ decay 
are energetic, while $q$ from the original $\tilde{q}$ decay is usually 
soft because of relatively small mass splitting between the squark and gluino. 
For $\tilde{q}_L$, another decay channel 
\begin{equation}
\tilde{q}_L\rightarrow{q}' \tilde{W}^{\pm}
\rightarrow{q}' \tilde{t}_1{b}
\label{eq:P3.10leftsqaurkdecay}
\end{equation}
could be important (although suppressed by the weak coupling) 
if the heavy gluino is off-shell in process (\ref{eq:P3.10squarkdecay}).

For $\tilde{b}_L$, a particular decay channel 
\begin{equation}
\tilde{b}_L\rightarrow\tilde{t}_1{W}^{\pm}\rightarrow
\tilde{t}_1(l\nu/{j}j')
\label{eq:P3.10leftsbdecay}
\end{equation}
dominates when the process (\ref{eq:P3.10squarkdecay}) is suppressed by
the small mass splitting $m_{\tilde{b}_L}-m_{\tilde{g}}$. 

For sleptons, a decay similar to (\ref{eq:P3.10squarkdecay}) occurs through 
neutral Wino and Bino:
\begin{equation}
\tilde{l}\rightarrow{l}\tilde{W}^0/\tilde{B}^0\rightarrow{l}\tilde{t}_1{t}
\rightarrow{l}\tilde{t}_1{b}W^{\pm}\rightarrow{l}\tilde{t}_1{b}(l\nu/jj').
\label{eq:P3.10sleptondecay}
\end{equation}
However, $\tilde{l}_L$ can also decay via the lepton
analogue of (\ref{eq:P3.10leftsqaurkdecay}),
which is comparable with (\ref{eq:P3.10sleptondecay}) for 
$m_{\tilde{W}}<m_{\tilde{l}}$ and dominates if 
$m_{\tilde{W}}>m_{\tilde{l}}$.

The gluino can decay via 
$\tilde{g}\rightarrow{t}\tilde{t}_1,\ b\tilde{b}_L,\ q\tilde{q}_L,\ 
q\tilde{q}_R$, 
either on-shell or off-shell, with subsequent decays of the 
$t$, $\tilde{b}_L$, $\tilde{q}_L$ or $\tilde{q}_R$ as described above.   
The generic decay products would be 
$\tilde{t}_1$+$b$+($l\nu/jj'$)+soft jet/lepton. 
Decays through $b\tilde{b}_L$ could lead to 
$\tilde{t}_1$+($l\nu/jj'$)+soft $b$ if the $\tilde{b}_L$
decays via (\ref{eq:P3.10leftsbdecay}), or $\tilde{t}_1$+$b$+soft jet is also 
possible if $\tilde{q}_L$ decays via (\ref{eq:P3.10leftsqaurkdecay}). 

There are three {\it Dirac} neutralinos/charginos, in contrast with 
the usual MSSM with four Majorana neutralinos and two Dirac charginos. 
The mass eigenstates are usually a mixture of gauginos and Higgsinos, 
with a small mass splitting of roughly 12 GeV (18 GeV) between 
the lightest one and the two heavier neutralinos (charginos) \cite{BHN}.  
The dominant decay channel for charginos is
$\chi^{\pm}\rightarrow{b}\tilde{t}_1$, 
leading to a clean signal of (track or missing energy)+$b$ jet.
Neutralinos decay similarly to gluinos or via 
$\chi^0\rightarrow\chi^{\pm}W^{\mp}$ with subsequent decays of $\chi^{\pm}$ 
and $W^{\mp}$. 

The heavier stop $\tilde{t}_2$ with mass $m_{\tilde{t}_2}=1/R+m_t$ 
can decay via 
\begin{equation}
\tilde{t}_2\rightarrow\tilde{t}_1Z\rightarrow\tilde{t}_1(2l/2\nu/2j),\ \ 
\tilde{t}_2\rightarrow\tilde{t}_1h\rightarrow\tilde{t}_1b\bar{b},\ \ 
\tilde{t}_2\rightarrow\tilde{b}_LW^{\pm}\rightarrow\tilde{t}_12(l\nu/jj')
\ + \ [b\ {\rm if}\ \tilde{b}_L\ {\rm decays\ via}\ (\ref{eq:P3.10squarkdecay})].
\label{eq:P3.10stopdecay}
\end{equation}

The heavy Higgs, with mass $m_H\sim{2}/R$,  
decays through $t\bar{t}$ like a usual heavy SM 
Higgs, while $H\rightarrow{WW}$ is forbidden due to the 
non-conservation of the fifth dimensional momentum.  
The discovery of the KK tower of
SM particles (heavy quarks, leptons, gauge bosons and their mirror partners) 
would be strong evidence for the existence of TeV-scale extra dimensions. 
However, they are  unlikely to be pair-produced at a TeV-scale 
$e^+e^-$ collider. 
For a hadron collider, the cascade decay chain is complicated 
and it is hard 
to distinguish the signal from the large background.  The discussion of 
the phenomenology of the heavier states is therefore left out 
in the current work.

\section{Experimental capabilities}
\label{P3.10expt}
\subsection{Tevatron Run II}

A heavy, stable charged particle (CHAMP) such as the $\tplus$($\tminus$)
predicted by the model will move slowly through the detector after it has 
been produced, losing large amounts of energy via ionization 
as it travels. The CDF Run I search~\cite{run1} for these types of particles
was based on the measurement of this energy loss by the tracking
system. The search found no evidence for the production of CHAMPS, 
and lower mass limits of $\sim$200 GeV were set 
in the context of a heavy quark model~\cite{hqu}. The limit in the 
context of SUSY models with a long-lived stop is currently
being evaluated.

Both the CDF and D$\O$ detectors have been upgraded in 
preparation for Run II.  The CDF upgrade includes a 
time-of-flight (TOF) detector~\cite{cdftof}.
With the TOF one can measure directly that a CHAMP
moves more slowly than a lighter particle at the same momentum. 
Furthermore, TOF
can distinguish between heavy and light particles 
at higher momentum than is possible with energy-loss measurements.
The CDF CHAMP search at Run II will therefore 
be based on measurements made by the TOF system.  D$\O$ plans to
use $dE/dx$ at the trigger level and their muon system for TOF 
\cite{dzerochamp}.

\begin{figure}
\begin{center}
\begin{tabular}{cc}
\epsfig{file=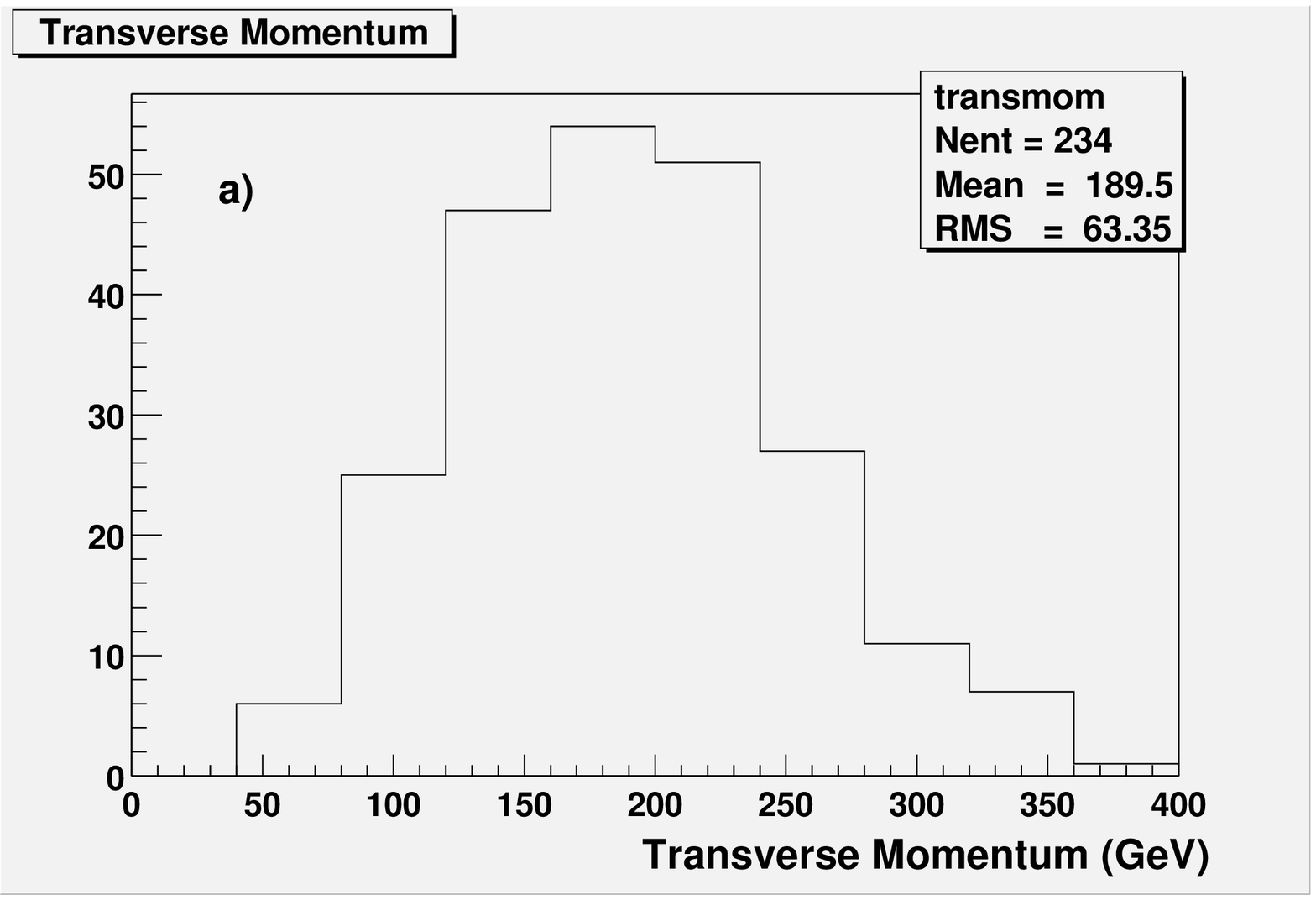,width=0.46\textwidth}
\epsfig{file=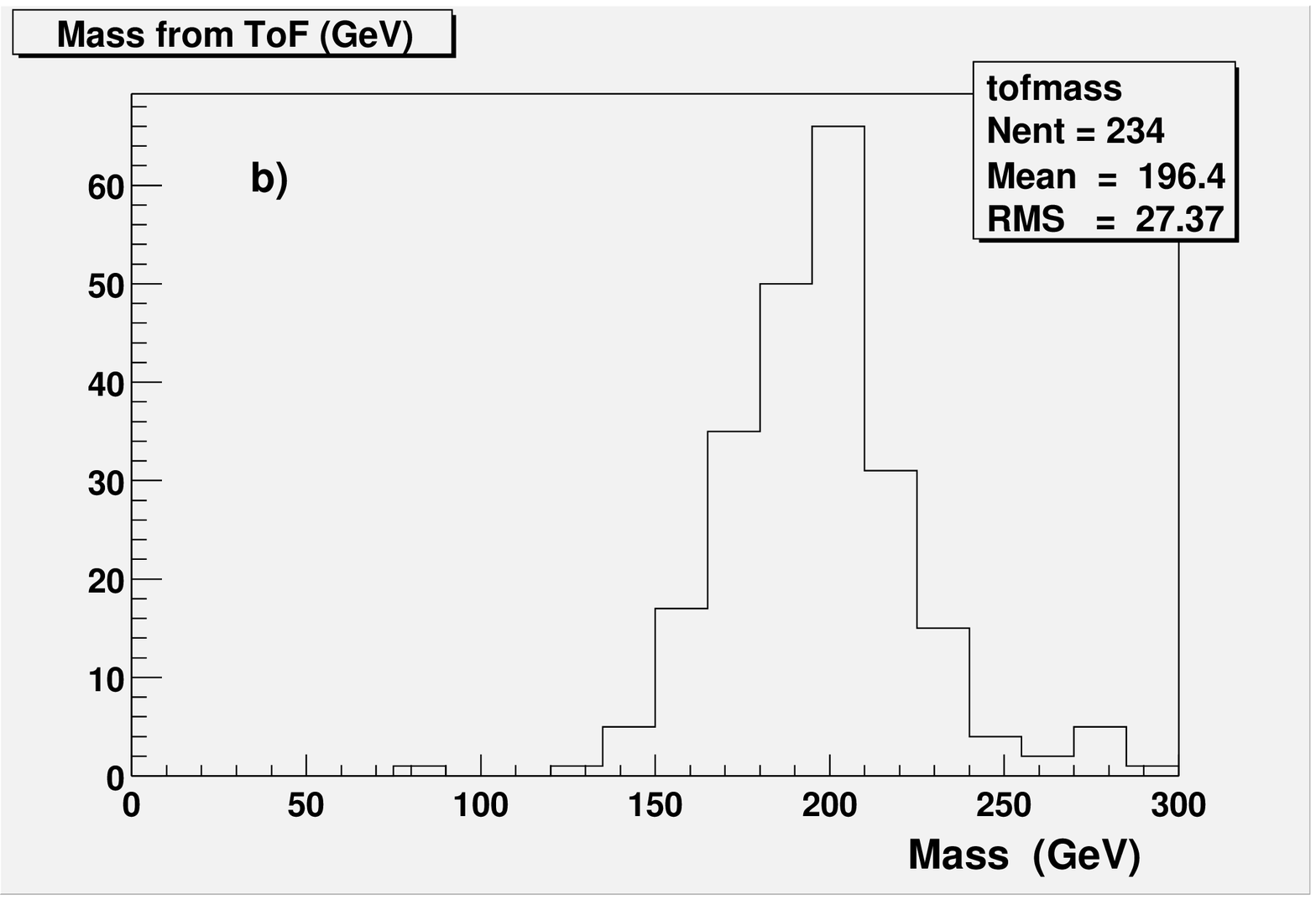,width=0.46\textwidth}
\end{tabular}
\end{center}
\caption{Monte Carlo distributions for a 200 GeV CHAMP
          at Run II of the Tevatron. (a) The transverse momentum after detector
	  simulation. (b) The mass after detector simulation. 
	  This mass has been 
          reconstructed by combining the momentum measurement with the
          TOF information. 
\label{P3.10quan}}
\end{figure}

Studies have been performed using Pythia~\cite{pyth} and the full CDF detector
simulation
to produce CHAMP Monte Carlo samples. 
The transverse momentum ($p_T$) distribution for a 
200 GeV CHAMP is shown in Figure~\ref{P3.10quan}a. The $p_T$ is
large, almost always greater than 50 GeV. The pair 
production of CHAMPs will thus leave two high-$p_T$
tracks in the detector as a signature. In order to take advantage of this
fact a high-$p_T$ two-track trigger has been developed to select events from
the Run II data. The trigger requires 2 isolated tracks
with $p_T$ $>$ 10 GeV. For a 200 GeV CHAMP, the trigger has
been found to be 37\% efficient and the rate from background (after
subtracting the rate from other triggers) has been estimated to be less than
10 nb at the nominal Run II luminosity of $10^{32}\; cm^{-2}s^{-1}$. 
Most signal events
which fail this trigger do so because one or both of the CHAMPs
in an event do not go through the fiducial tracking volume.
In order to increase the trigger efficiency, a high $p_T$ single track 
trigger is also being developed.

Events that are selected by the triggers will be analyzed to
identify particles with a large TOF. This information can then be used with
the momentum obtained from the tracking system to measure the 
mass of the particle in question.  Figure~\ref{P3.10quan}b 
shows this measured mass for a 200 GeV CHAMP 
Monte Carlo sample and demonstrates that such 
particles can be reconstructed with a resolution of about 25 GeV. 
The dominant effect is the uncertainty inherent in 
measuring the transverse momentum for tracks with 
very little curvature.

In this model, the cross section for the pair production of light stops is 
about 0.6 pb
\cite{tevst}, leading to about 600 events per year at 
nominal luminosity.\footnote{The cross section is twice 
of that in Ref.~\cite{tevst}
because of the multiplicity of the degenerate stops.}
Triggerable events would have one or two high-$p_T$ 
isolated tracks; events with
two neutral $T^0$ from direct production might be 
impossible to trigger on 
unless other appreciable activity is present.
The scalar nature of the top squark could be confirmed by the 
angular distribution of the differential cross section 
${\rm d}\sigma/{\rm d}\cos\theta$. 
The measurement of the light stop mass in the predicted range 
with twice the standard cross section
would be compelling evidence for this model. 

The cross section for 
$\tilde{q}\tilde{q}$, $\tilde{q}\tilde{g}$,
$\tilde{g}\tilde{g}$ production ranges from $0.1$ to 1 pb in this model
\cite{tevsquark}.  The final decay 
products always include two $\tilde{t}_1$, which can be triggered on as 
discussed above.  
In addition, there will be other objects that can be used to 
distinguish these events from direct stop pair production.  These include
two energetic $b$ jets (except for the case of 
$\tilde{b}_L$ which decay via Eq.~(\ref{eq:P3.10leftsbdecay})), possible energetic 
leptons, $\slash\hspace{-0.1 in}{E}_T$ (from neutrinos), 
and jets coming from decays of the intermediate $W$.  
Alternative triggers based on these objects should
be useful in selecting events containing two neutral $T^0$, although
distinguishing the signal from background requires detailed study.
Finally, the production of sleptons, gauginos and 
higher mass states is less promising at the Tevatron in Run II.

\subsection{LHC}

Given the factor of seven increase in center of mass energy over
Tevatron Run II, the LHC experiments will substantially extend
the search for phenomena predicted by this model.  In particular, the
heavier mass states should be accessible.  If the light Higgs is not
discovered at the Tevatron it will be uncovered at the LHC in the
usual channels.  The heavy
Higgs discovery should also be possible, as in the SM case, provided enough
integrated luminosity is achieved \cite{lhc-higgs}.  
Nevertheless, prospects for identifying either of these Higgs as 
unique to this model requires more study.  One strategy would search
for the heavy Higgs decaying via some of the predicted SUSY states.

Light stop pairs can be produced copiously with a cross section of 
$\sim$ 150 pb \cite{lhcst}.  
As at the Tevatron, the charged stop hadrons should give 
rise to unmistakable detector signatures at the LHC, provided these events
are selected by online triggers.  For example, muon triggers that require
low $dE/dx$ in the tracking chambers might reject these events and
thus should be avoided.
Sensitivities to stop can be estimated from a CMS collaboration projection 
for a NLSP stau in GMSB models \cite{cms-stau}.
For a long-lived stau, the authors utilize the drift tubes in the muon barrel
as a time-of-flight detector; the available timing
window for this approach requires $1/\beta < 2$.  
A clear signal is observed: the efficiency grows with
stau mass and the backgrounds considered are easily suppressed.  For
$M(\tilde{\tau})=636\;$ GeV the efficiency is 26\%.
The $T^{\pm}$ in this model could be
identified in a similar fashion provided that they are in a charged
state when traversing the muon chamber.

Squarks and gluinos can also be produced copiously and the charged
tracks can be used for triggering, as before.  
In addition, events with neutral stop mesons could create a large $\missinget$ 
event signature and be selected by triggers requiring an associated
high-$p_T$ lepton or jet.
Even for the heavier stop states, hundreds of events per year could
be observed, given the clean signal of $\tilde{t}_2$ decay via 
Eq.~(\ref{eq:P3.10stopdecay}).  The measurement of the mass difference between
$\tilde{t}_2$ and $\tilde{t}_1$ (predicted to be $2m_t$) would be another 
check of the model. 
It should be noted that the prediction of doubled cross sections 
due to degeneracies of the various 
SUSY states would provide a means for differentiating these 
signals from typical SUSY models. 

\subsection{Linear Collider}

A high-energy electron-positron Linear Collider (LC) running at
center-of-mass energies of up to 1 TeV, would permit a number
of measurements which would be essential to the confirmation
of this model. With a mass predicted to be less than 250 GeV,
pair production of the light stop would be copious even at
a 500 GeV LC. The factor-of-two enhancement of the stop production
cross section (due to the presence of the degenerate mirror state)
could be measured to a statistical and systematic
precision of order 1\%. At higher energies, individual thresholds for the full
spectrum of particles at the scale of $1/R$ could be detected. The
resulting picture -- a light stop plus a nearly-degenerate array of sfermions,
Higgsino states, and gauginos, with each sfermion exhibiting a 
precise factor-of-two production
rate enhancement -- would represent an unambiguous signature for this model.
Another unique characteristic  -- the Dirac nature
of the neutralinos and charginos -- could be confirmed by studying the
helicity dependence of their production cross-sections.
The associate production of $\tilde{t}_1\tilde{t}_2$ is possible at a TeV 
LC via 
$s$-channel $Z$-exchange.  

Several other checks of the model, which are not possible at a hadron machine,
could be performed at a Linear Collider. As mentioned above, the $O(10\%)$
deviation from the SM top Yukawa coupling could be measured to 1-2$\sigma$
at a 1 TeV LC~\cite{tthlc}.
The chiral composition of the stop mass eigenstates could
also be confirmed by exploiting the beam polarization. All in all,
the ability to individually detect and precisely measure the properties
of each of the 1/R-scale states,
including those (such as sleptons) which are
not accessible to hadron machines, makes the complementary information provided
by the LC an essential component of the confirmation of the model's
characteristic phenomenology.

\section{Summary}
We have described a model for supersymmetry breaking in 5 dimensions and 
developed its phenomenology and the capabilities for discovery and 
measurement at various colliders.
Discovery of the predicted states is possible 
at the LHC and perhaps at the Tevatron in Run II.
The degeneracy of each state leading to factor-of-two enhanced 
sfermion cross sections could also be measured.
However, detailed measurements at a high-energy
LC will be necessary to confirm this model.
The discovery of the Kaluza-Klein tower of
SM particles (heavy quarks, leptons, gauge bosons and their mirror partners) 
would constitute strong evidence for
the existence of TeV-scale extra dimensions. 

%
%

%
%




\end{document}